\documentclass{PoS}

\title{Fermi/LAT Gamma Ray Burst emission models and jet properties}

\ShortTitle{Fermi/LAT GRBs}

\author{\speaker{Gabriele Ghisellini}\\ 
        INAF -- Osservatorio Astronomico di Brera \\
        E-mail: \email{gabriele.ghisellini@brera.inaf.it}}

\author{Giancarlo Ghirlanda\\
        INAF -- Osservatorio Astronomico di Brera \\
        E-mail: \email{giancarlo.ghirlanda@brera.inaf.it }}

\abstract{
The GeV emission of Gamma Ray Bursts, first detected by the 
Energetic Gamma--Ray Experiment Telescope (EGRET) onboard the
{\it Compton Gamma Ray Observatory} in an handful of bursts,
is now an established property of roughly the 10\% of all (both short and long)
bursts, thanks to the  {\it Fermi}/Large Area Telescope (LAT) observations.
GRB 090510, a short burst, is particularly interesting
because the good timing allows to derive a severe limit to theories
of quantum gravity predicting an energy dependent delay of the arrival of photons.
Up to now there have been a dozen bursts detected in the 0.1--30 GeV band,
and despite the small number, we start to see some common properties:
(i) the duration is often longer than the duration of the softer emission
detected by the Gamma Burst Monitor (GBM) onboard {\it Fermi};
(ii) the spectrum is consistent with
$F_{\nu}\propto \nu^{-1}$ with no strong spectral evolution;
(iii) for the brightest bursts, the flux detected by the LAT decays as a power law
with a typical slope: $t^{-1.5}$;
iv) the peak energy of the GBM emission is large, exceeding 500 keV in the
rest frame of the burst.
These common properties suggest a similar dominant process for the origin
of the GeV flux. 
We propose that it is afterglow synchrotron emission shortly 
following the start of the prompt phase emission seen at smaller frequencies.
The steep decay slope suggests that the fireball emits in the
radiative regime, i.e. all dissipated energy is radiated away.
The large peak energy of the GBM flux suggests that electron--positron pairs 
might play a crucial role for the setting of the radiative regime.
The rapid onset, but with some delay, of the GeV flux with respect to the 
GBM one suggests that the bulk Lorentz
factor $\Gamma$ of these bursts is large, of the order of 1000.
Therefore the relatively small fraction of bursts detected at high energies might
correspond to the fraction of bursts having the largest $\Gamma$.
If the emission occurs in the radiative regime we can start to understand 
why the observed X--ray and optical afterglow energetics are much smaller
than the energetics emitted during the prompt phase, despite the
fact that the collision with the external medium should be more
efficient than internal shocks in producing the radiation we see.
 }

\FullConference{The Extreme sky: Sampling the Universe above 10 keV - extremesky2009,\\
		October 13-17, 2009\\
		Otranto (Lecce) Italy}

\begin{document}

\section{Introduction}

Since the detection, by EGRET, of an handful of Gamma Ray Bursts (GRBs) 
above 100 MeV, we have been left with the question: 
does this emission belong to the prompt phase or is it afterglow
emission produced by the fireball colliding with the circum--burst
medium? Or has it still another origin?
A puzzling feature of the EGRET high energy emission
was that it was long lasting, yet it started during the prompt
phase as seen by the Burst Alert and Transient Experiment (BATSE, 30 keV -- 1 MeV)
onboard {\it CGRO}.

Now, with {\it Fermi}/LAT \cite{atwood2009}, we can start 
to address these issues.
It revealed, up to October 2009, 12 GRBs above 100 MeV. 
This (still small) sample of bursts starts to show 
some regularities, even if the situation is rather complex.
For instance, GRB 080916C \cite{abdo2009a}, 
has a 8 keV -- 10 GeV spectrum that can be described by the same Band
function (i.e. two smoothly connected power laws), indicating
that the LAT flux has the same origin of the low energy flux.
On the other hand, the level of the LAT flux and its spectrum are similar to 
the emission from forward shocks, leading \cite{kumar2009} to prefer 
the ``standard afterglow" interpretation [see also \cite{razzaque2009}
for an hadronic model; \cite{zhang2009} for a magnetically 
dominated fireball model and \cite{zou2009} for a synchrotron self--Compton 
(SSC) origin].
In other bursts the LAT flux has a spectrum that is harder than
the extrapolation from lower energies, as in the short bursts GRB 090510,
leading  \cite{abdo2009b} to propose a SSC interpretation
(but see \cite{ghirlanda2010}; \cite{gao2009}; \cite{depasquale2009}
for alternative views).
Due to the optimal timing this burst is ideal to set constraints
\cite{amelino1998}, \cite{abdo2009b}, \cite{ghirlanda2010} on theories
of quantum gravity predicting the violations of  
Lorentz invariance.

Considering the ensemble of bursts, a consistent scenario seems to emerge: 
the LAT spectra are often inconsistent with the extrapolation of the GBM spectra
(except two cases) and the light curves can be described
by a power law decay in time, i.e. $F_{\rm LAT}\propto t^{-\alpha}$,
with a slope close to $\alpha=1.5$.
In the brightest cases also the rising part is visible, and
is consistent with $F_{\rm LAT}\propto t^2$.
These are indications of the afterglow nature of the LAT emission.
GRBs with a flux decaying as $F_{\rm LAT}\propto t^{-1.5}$, and with a 
spectral slope around unity [i.e. $F(\nu)\propto \nu^{-1}$]
could be emitting in the radiative regime of a forward shock,
running in a medium enriched by electron--positron pairs.

\begin{figure}
\includegraphics[width=1\textwidth]{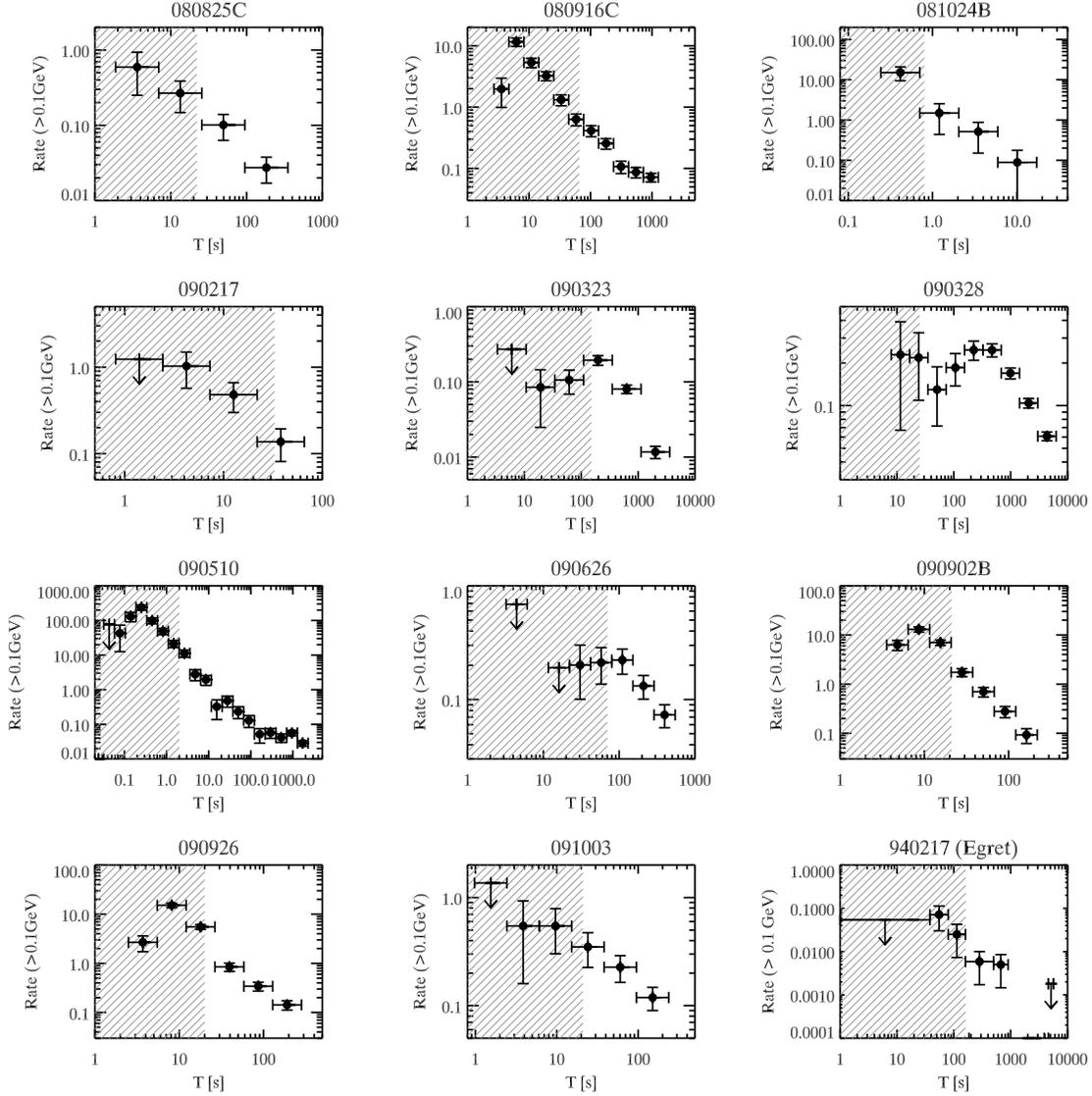}
\caption{
Light curves of the 11 GRBs detected by LAT 
plus GRB 940217, as detected by EGRET (bottom right panel).
The hatched region represents the duration
($T_{90}$) of the emission detected by the GBM in the 8 keV--40 MeV energy range
(for GRB 940217 it refers to the emission detected by BATSE).
Times are in the observer frame for all bursts and arrows represent 
2$\sigma$ upper limits. From \cite{ghisellini2010}.
}
\label{fig1}
\end{figure}

\section{LAT light curves}

Fig. \ref{fig1} shows the $>0.1$ GeV light curves.
The hatched regions indicate the $T_{90}$ duration of the GBM light curve. 
In 9/11 events there is 
a peak in the LAT light curve and the latter has a duration much longer than
the duration of the GBM light curve. 
After the peak, the light curves of different GRBs show a similar temporal decay. 
In a few cases a rising of the light curve as 
$t^2$ is seen before the peak \cite{ghirlanda2010}. 
The three faintest GRBs (GRB 090323, GRB 090328 and GRB 090626)
have light--curves that appear much flatter than the other ones
(note the different scale of their $y$--axis) and we cannot exclude
that the background, in this cases, plays some role.
The bottom right panel shows the light--curve of GRB 940217
as detected by EGRET \cite{hurley1994}, selecting photons above 100 MeV.
As can be seen, also this burst show a similar decaying light curve.

\subsection{General properties of the LAT bursts}

From the analysis of all GRBs detected by the LAT we have found these
properties \cite{ghisellini2010}:
\vskip 0.2 cm
\noindent
{\bf No spectral evolution --} 
The time resolved spectral results of individual bursts  
show no evidence of strong spectral evolution of the LAT spectral index.

\vskip 0.2 cm
\noindent
{\bf LAT and GBM spectral slopes are often different --}
Only in two bursts, GRB 080916C (\cite{abdo2009a} and GRB 090926 
the high energy (steep) slope of the GBM data is consistent with 
the slope of the LAT data.
For all the rest, the LAT slope is softer (harder) than the low (high) energy
slope of the GBM data.

\begin{figure}
\includegraphics[width=0.54\textwidth]{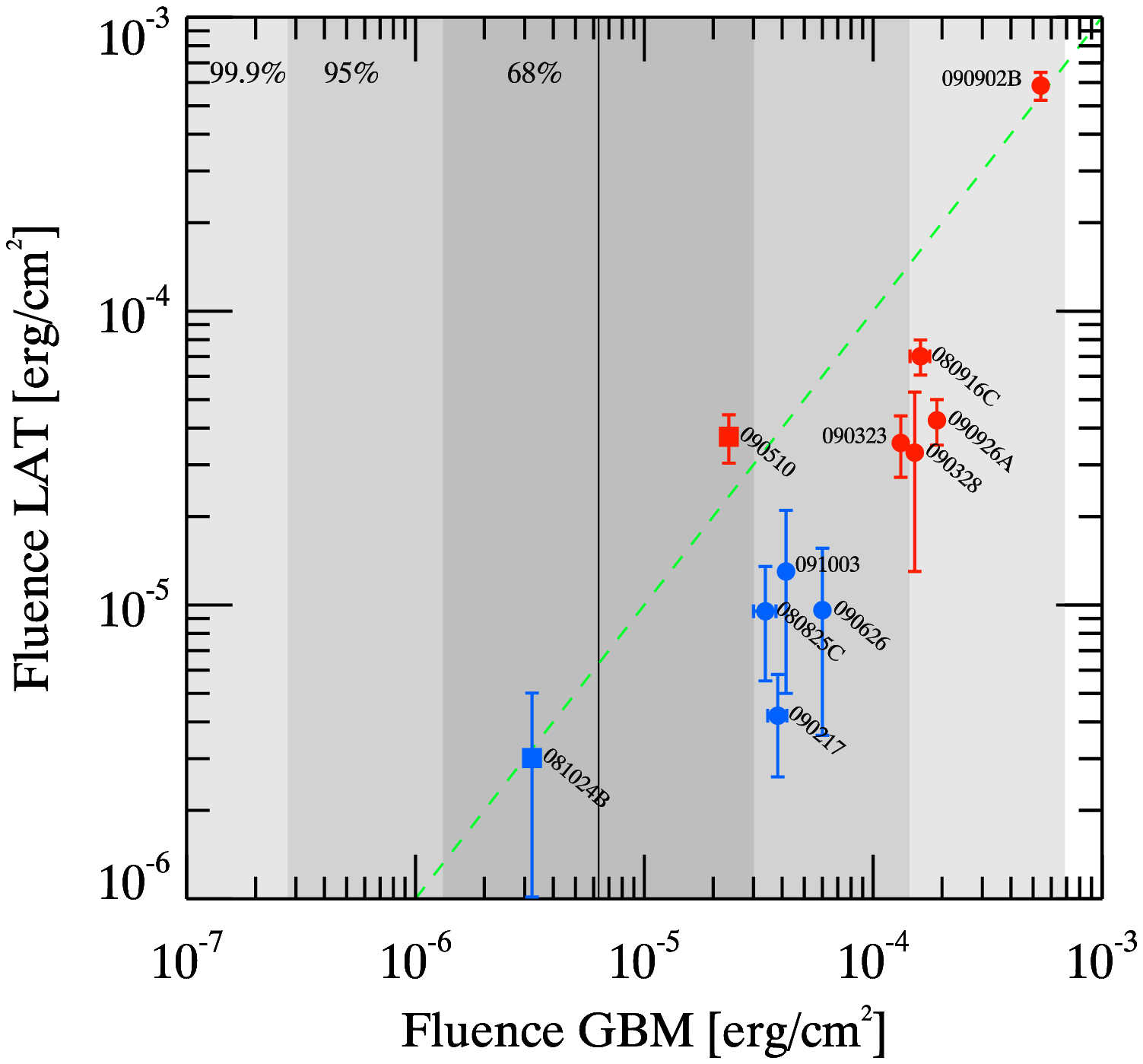}
\includegraphics[width=0.48\textwidth]{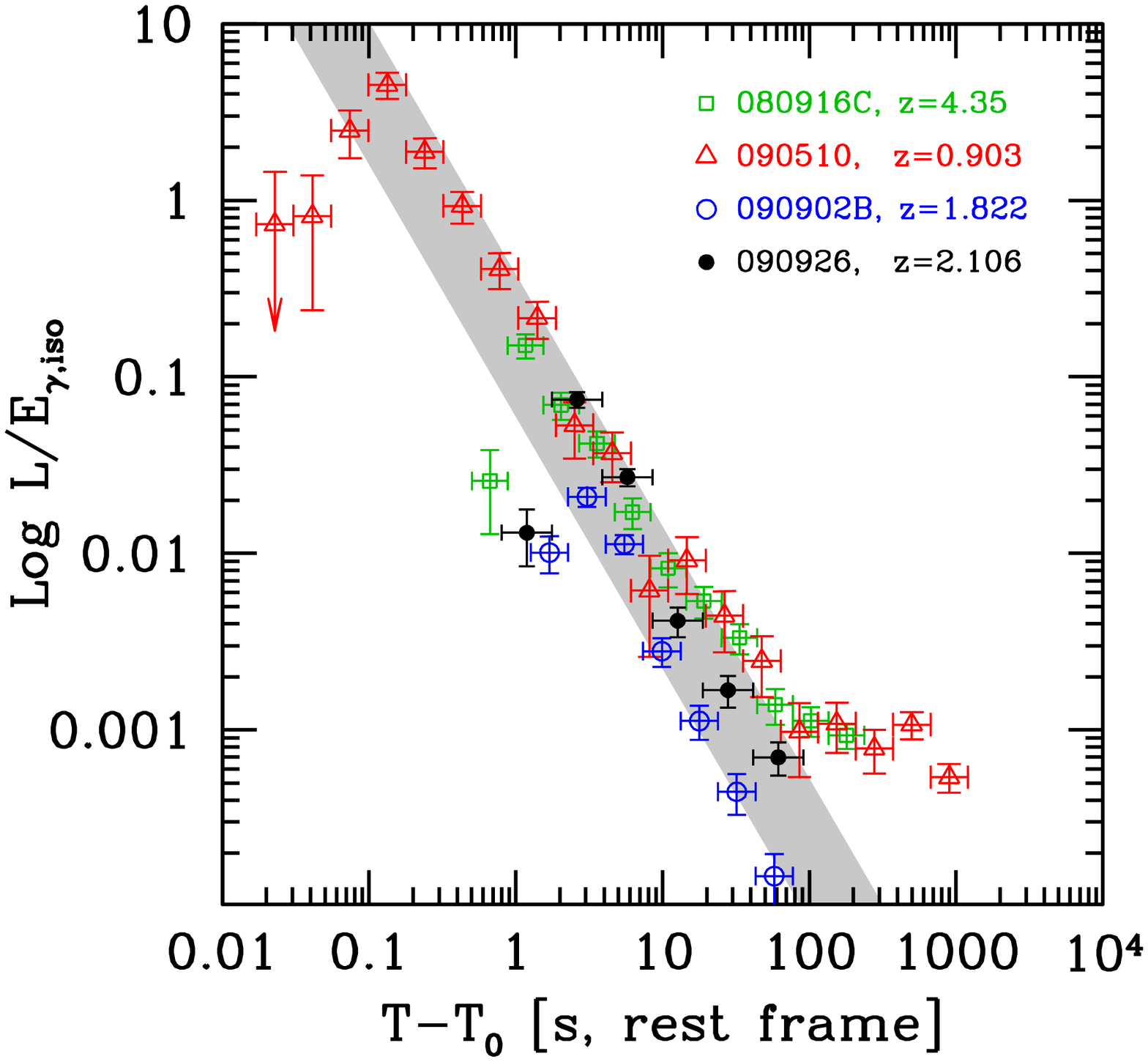}\\
\caption{
Left: the [0.1--100 GeV] LAT fluence  as a function of the 
[8 keV -- 10 MeV] GBM ones.
Filled squares and circles correspond to short and long GRBs, respectively.
The shaded areas indicate the 1--2--3 $\sigma$ values
of the distribution of GBM fluences of the 121 GRB 
with $E_{\rm peak}$ (as of Oct. 2009).
Right: Light curves of the 4 brightest GRBs with redshift,
normalised to the total energetics of the GBM data.
The luminosities are 
integrated in the 100 MeV--100 GeV energy range at the source rest frame. 
The time is in the rest frame of the sources. 
The yellow stripe indicates a slope t$^{-10/7}$.  From \cite{ghisellini2010}
}
\label{fig2}
\end{figure}

\vskip 0.2 cm
\noindent
{\bf LAT fluences are smaller than GBM ones --} 
Fig. \ref{fig2} (left) shows that the majority of bursts have LAT fluences
smaller than the GBM ones, except for
the two short bursts GRB 081024B and GRB 090510
and for GRB 090902B (for which they af the same order).

\vskip 0.2 cm
\noindent
{\bf Common decay for the brightest LAT bursts --}
Fig. \ref{fig2} (right) shows the light curves of the 
4 brightest GRBs with
redshift, once the 0.1--100 GeV luminosity is divided by 
the energetics $E_{\rm \gamma, iso}$ of the flux detected by the GBM.
The shaded stripe with slope $t^{-10/7}$ is shown for comparison.
These four GRBs are all consistent,
within the errors, with the same decay, both in slope and in normalisation.
Note that GRB 090510, a short burst, behaves similarly to the other 3 bursts,
that belong to the long class, but its light--curve begins much earlier.


\vskip 0.2 cm
These properties are just what expected by the 
external shock scenario giving rise to the afterglow.
We therefore suggest that the high energy emission of the
GRBs detected by the LAT has an afterglow origin.
The high energy emission can overlap in time with the 
prompt GBM phase if $\Gamma$ is large, making the fireball 
to decelerate earlier and shortening the observed times
by the Doppler effect.
What is at odd with respect to the standard afterglow scenario
is the relatively steep slope of the flux decay, $F(t)\propto t^{-1.5}$,
instead of the more comfortable $F(t)\propto t^{-1}$ 
expected for the decay of the bolometric flux in the fast cooling 
and adiabatic regime.

\section{Radiative fireballs}

The relatively steep decay of the LAT light curves is very close to
what expected if the fireball emits in the radiative regime.
In this case the bolometric flux in fast cooling decays as
$F(t)\propto t^{-10/7}\sim t^{-1.43}$ \cite{sari1998}, \cite{ghisellini2010}.
The spectral slopes of the LAT emission are close to unity (in energy), ensuring
that we are seeing emission close to the peak in $\nu F_\nu$, 
so the LAT fluxes are a reasonably good proxy for the bolometric ones.
A radiative regime occurs when all the energy dissipated by the forward shock
is radiated away. 
There is no equipartition between the magnetic field, 
the protons and the electrons, all the energy goes to electrons, 
the only ones that can radiate.
A possibility is that there is an efficient exchange of energy between protons
and electrons, which is however difficult to envisage, because 
the relevant number density of particles is small, and two--body interactions
are disfavoured.
But there is another possibility: pair enrichment of the circum--burst medium
by the prompt radiation, as envisaged by \cite{belo2002}.
The optical depth of the medium is very tiny, and the vast majority
of the prompt photons passes undisturbed.
However, each electron of the circum--burst medium (within  $\sim 10^{-15}$--$10^{17}$ cm
from the burst) does scatter a large quantity of prompt photons
(there are few electrons and many photons).
The scattered photons change direction of propagation and become easy and 
efficient targets 
for the $\gamma$--$\gamma\to e^\pm$ process with those photons of the prompt
with energy larger than $\sim$1 MeV.
These pairs can even contribute to the process, scattering even more prompt photons.
During each scattering, some momentum is deposited to the electrons, and eventually 
the medium move forward relativistically, quenching the process (see \cite{belo2002}
for details).

If the medium has been enriched by pairs, then
it is conceivable that the shock acceleration process will give more energy
to leptons rather than to protons and magnetic field, simply because leptons  
are now much more numerous. 
The pair--enrichment process could then imply a radiative regime.
The radiative phase should end either because the cooling becomes inefficient
(i.e. when the slow cooling phase starts), or because we run out of pairs,
i.e. the numbers of pairs produced at a given distance from the burst becomes
less than the original electrons.

Note a key ingredient of this scenario: to produce pairs efficiently, 
the prompt emission should have a sufficient number of photons above threshold,
i.e. above 511 keV.

\begin{figure}
\begin{center}
\includegraphics[width=0.35\textwidth]{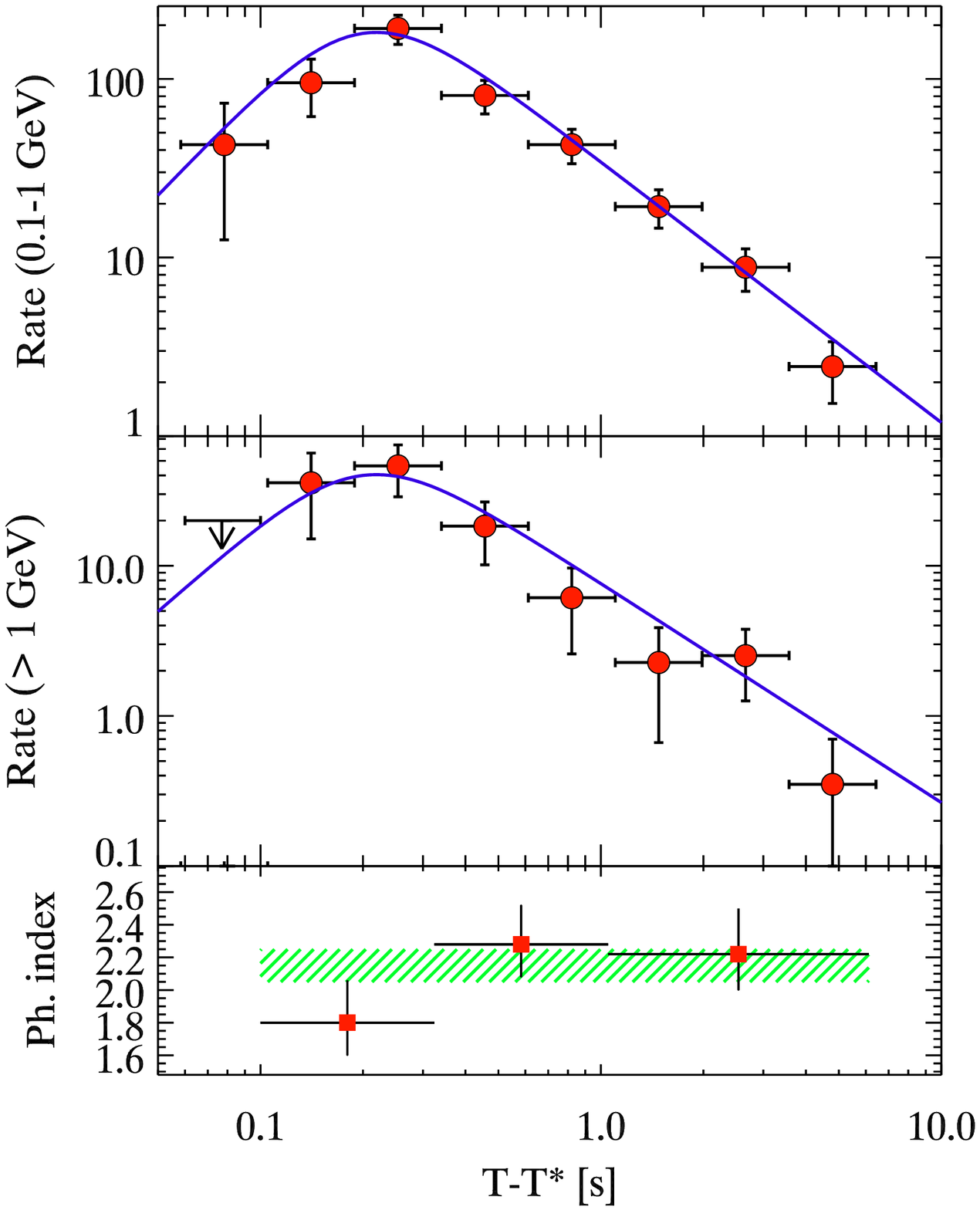} 
\includegraphics[width=0.52\textwidth]{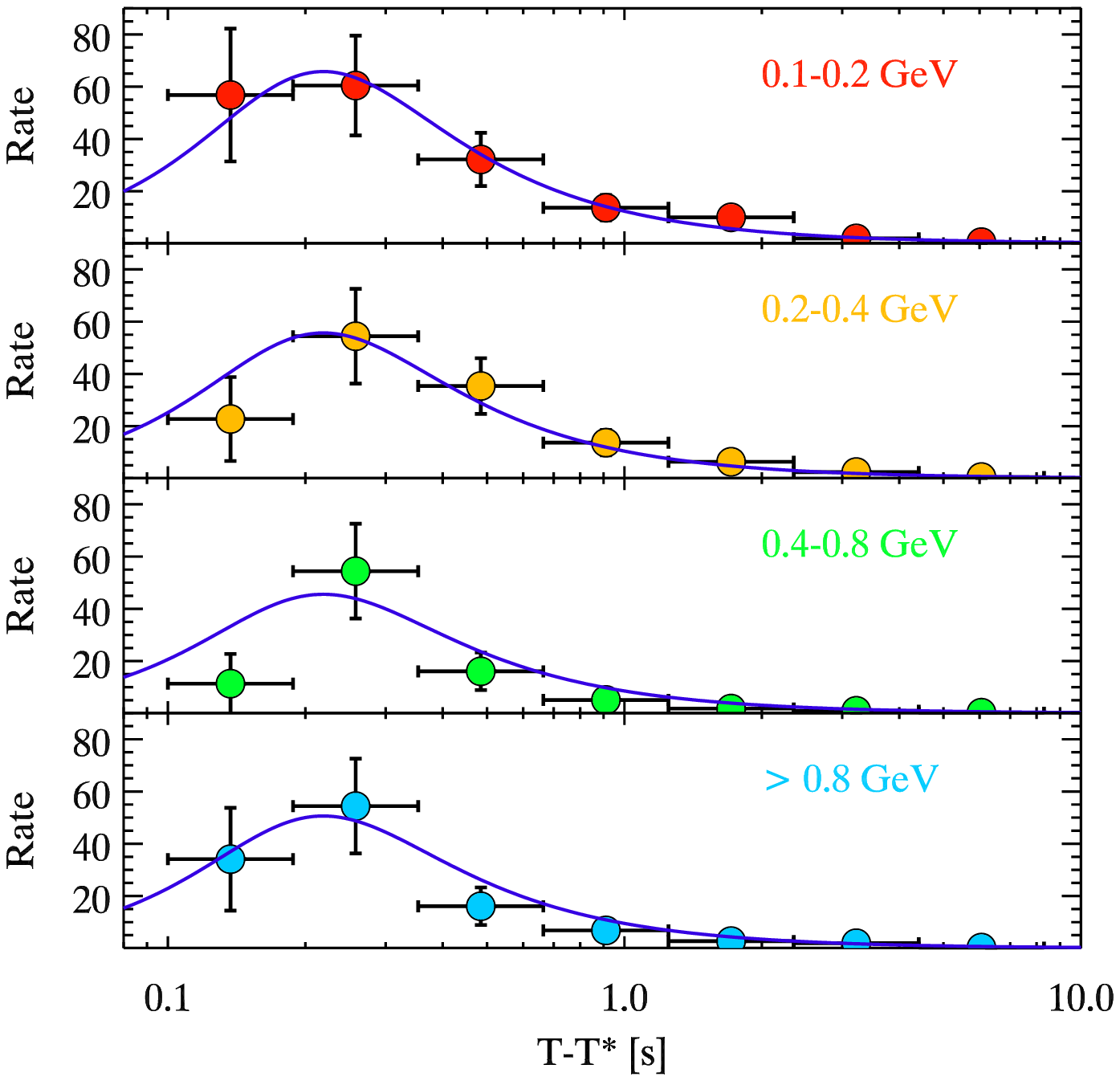} \\
\end{center} 
\caption{
Left:
{\it Fermi}--LAT light curve of GRB 090510 between 0.1 and 1 GeV 
and above 1 GeV
(top and middle panels, respectively) in the first 10 seconds. 
The times are scaled to $T^{*}$=0.6 s. 
The solid line is the fit of the light curve $>$0.1 GeV. 
The bottom panel shows the photon spectral index
of the LAT spectra for the time--integrated spectrum (hatched region) and for 
three time resolved spectra (squares).
Right:
light curve of GRB 090510 in four energy channels 
(from top to bottom): 0.1--0.2 GeV, 0.2--0.4 GeV, 0.4--0.8 GeV, $>$0.8 GeV. 
From \cite{ghirlanda2010}.
}
\label{fig3}
\end{figure}

\section{The short burst GRB 090510: limits to violations of the 
Lorentz invariance}

GRB 090510 is a short/hard burst at redshift $z$=0.903 \cite{rau2009}  
detected by {\it Fermi} \cite{guiriec2009}, AGILE \cite{longo2009}, 
{\it Swift} \cite{hoversten2009}, 
{\it Konus-Wind} \cite{golenetskii2009} and {\it Suzaku} \cite{ohmori2009}.
The GBM triggered on a precursor while the main emission episode in the 
8 keV--40 MeV energy range 
starts $\sim$0.5 s after trigger and lasts up to $\sim$1 s. 
The emission observed by the LAT starts 0.65 s after the trigger and lasts $\sim$ 200 s.  
The joint GBM--LAT spectral analysis showed the presence of two components. 
{\it Fermi}--LAT detected a 31$\pm3$ GeV photon delayed by 0.829 s with respect 
to the trigger \cite{abdo2009b}.
Therefore the $\sim$MeV emission component is followed 
by a longer lasting high energy emission detected above 100 MeV. 
Both the AGILE and {\it Fermi}\ spectra suggest that this component is not 
the extrapolation of the soft $\sim$MeV spectrum to the GeV range. 
\cite{abdo2009b} interpret the $\sim$MeV flux as synchrotron radiation and the 
LAT flux as its synchrotron SSC emission. 
The detection of a 30 GeV photon sets a lower limit on the bulk Lorentz 
factor of the fireball $\Gamma>1000$, based on the compactness 
argument \cite{abdo2009b}. 
The 30 GeV photon arrives 0.829 s after the trigger (set by the precursor) and 
0.3 s after the beginning of the GBM main pulse. 
These delays allowed \cite{abdo2009b}  to put 
limits on the violation of the Lorentz invariance. 

We \cite{ghirlanda2010} proposed a different interpretation:
if $\Gamma>1000$ the fireball should start to decelerate and produce  
a luminous afterglow rather early (e.g. \cite{piran2004}), 
even at the sub--second timescale.
The light curves of the LAT flux for different energy intervals
(Fig. \ref{fig2}) all show the same rise and decay behaviour, and can be
fitted with the same law \cite{ghirlanda2010}.
Based on this and on the LAT spectra
we suggested that the flux detected by the LAT
is afterglow synchrotron emission of the forward external shock.
This radiation is produced at later times than the GBM flux,
and it is not delayed by possible violations of the Lorentz invariance.
It is then possible to derive a stringent lower limit on the corresponding 
quantum--gravity mass that must be greater than 
5 Planck masses \cite{ghirlanda2010}.


\section{Conclusions}


\begin{itemize}
\item
The compactness argument implies that if the $\sim$GeV emission is 
cospatial with the keV--MeV radiation, the bulk Lorentz factor must be large.

\item
But if $\Gamma$ is large, then the onset of the afterglow is rapid, even
fractions of a second after trigger.
Therefore a robust statement is that bursts detected by the LAT
have a large $\Gamma$ (i.e. $>1000$) in any case.

\item
In external shocks, a large $\Gamma$ implies large electron energies and 
magnetic fields, therefore large synchrotron frequencies.
Inverse Compton frequencies (both by SSC and by scattering ambient radiation)
are so large that are typically outside the LAT energy range. 
Furthermore, Compton scatterings likely occur in the low efficient Klein Nishina regime.
Instead, there is no strong objection to the fact that the LAT emission is 
afterglow synchrotron radiation.

\item 
If the MeV and GeV photons belong to the prompt and afterglow phases,
respectively, the (short) delay between the start of the two phases can be very 
easily understood. 
It has noting to do with violations of the Lorentz invariance, that can be tested
more reliably by considering only the high energy emission, that belongs
to the same component.

\item 
The LAT light curves and spectra resemble very closely 
what we expect from an afterglow origin of the emission.

\item
The best (i.e. brightest) bursts show a remarkable similar LAT light curve.
All decay as $F(t)\propto t^{-1.5}$, suggesting afterglow emission in the
radiative regime.

\item
The fact that the prompt GBM spectrum of these bursts peaks at or above 
$\sim$500 keV suggests that the radiative regime could be set by the transformation
of even a tiny amount (one in a billion is sufficient) of prompt $\sim$MeV photons
into electron--positron pairs.
These pairs enrich the circum--bursts medium by leptons, helping the 
forward shock to give most of its energy to leptons, rather than to protons 
or to the magnetic field.

\item
There is an immediate test to this idea: X--ray flashes should 
not have a radiative phase in their afterglows.

\item
A rapid onset of the afterglow implies a large initial luminosity,
and therefore a better chance to be visible by the LAT.
Thus the LAT bursts might be the ones with the largest $\Gamma$,
i.e. the ones for which the deceleration of the fireball occurs earlier.
The fraction of LAT--detected bursts over the ones detected by the GBM
is roughly 10\%, once accounting for the different fields of view.
This fraction may then be the fraction of high--$\Gamma$ bursts.

\item
Bursts with smaller $\Gamma$ may also produce {\it energetic} $\gamma$--ray afterglows,
but with {\it smaller luminosities}, since the peak of their emission occurs later.
LAT could may then detect a relatively nearby burst with a $\gamma$--ray
flux much more delayed (i.e. tens of seconds) with respect to the GBM detection.

\item
Since external shocks are more efficient 
than internal ones to dissipate the kinetic energy of the fireball, 
we expect the energetics (or the fluences) of the afterglows 
to be greater than the energetics (or fluences) of the prompt phase.
We have seen the opposite so far (e.g. \cite{zhang2007}, \cite{willingale2007}),
but if most bursts generously emit at high energies during the early phases
of their afterglows, we have to revise the energetic budget of the afterglow.

\end{itemize}

\end{document}